\documentclass[pra,twocolumn,superscriptaddress,floatfix]{revtex4}
\usepackage{amsmath}
\usepackage{amssymb}
\usepackage{amsthm}
\usepackage{longtable}
\usepackage{rotating}
\usepackage{url}

\usepackage{bm}

\newcommand{\A}{\mathrm{A}}
\newcommand{\B}{\mathrm{B}}
\newcommand{\I}{\mathrm{I}}

\newcommand{\trans}{\mathrm{T}}
\newcommand{\vct}{\bm}

\newcommand{\CutP}{\mathrm{CUT}^\square}
\DeclareMathOperator{\tr}{tr}
\newcommand{\CC}{\mathbb{C}}

\theoremstyle{plain}

\newtheorem{cnjc}{Conjecture}
\theoremstyle{definition}

\begin{document}
\title{Bell inequalities stronger than the CHSH inequality \\
  for 3-level isotropic states}
\date{January~25, 2006}
\author{Tsuyoshi Ito}
\email{tsuyoshi@is.s.u-tokyo.ac.jp}
\affiliation{Department of Computer Science, University of Tokyo}
\author{Hiroshi Imai}
\email{imai@is.s.u-tokyo.ac.jp}
\affiliation{Department of Computer Science, University of Tokyo}
\affiliation{ERATO-SORST Quantum Computation and Information Project,
  Japan Science and Technology Agency}
\author{David Avis}
\email{avis@cs.mcgill.ca}
\affiliation{School of Computer Science, McGill University}


\bibliographystyle{apsrev}

\begin{abstract}
  We show that some two-party Bell inequalities with two-valued
  observables are stronger than the CHSH inequality for $3\otimes3$
  isotropic states in the sense that they are violated by some
  isotropic states in the $3\otimes3$ system that do not violate the
  CHSH inequality.
  These Bell inequalities are obtained by applying triangular
  elimination to the list of known facet inequalities of the cut
  polytope on nine points.
  This gives a partial solution to an open problem posed by
  Collins and Gisin.
  The results of numerical optimization suggest that they are
  candidates for being stronger than the $\I_{3322}$ Bell inequality
  for $3\otimes3$ isotropic states.
  On the other hand, we found no Bell inequalities stronger than the
  CHSH inequality for $2\otimes2$ isotropic states.
  In addition, we illustrate an inclusion relation among some Bell
  inequalities derived by triangular elimination.
\end{abstract}


\maketitle

\section{Introduction}

Bell inequalities and their violation are an important topic in
quantum theory~\cite{WerWol-QIC01,KruWer-0504166}.
Pitowsky~\cite{Pit-JMP86,Pit-MP91} introduced convex polytopes called
\emph{correlation polytopes} which represent the set of possible
results of various correlation experiments.
A Bell inequality is an inequality valid for a certain correlation
polytope.
The correlation experiments we consider in this paper are those between
two parties, where one party has $m_\A$ choices of two-valued
measurements and the other party has $m_\B$ choices.
The Clauser-Horne-Shimony-Holt inequality~\cite{ClaHorShiHol-PRL69} is
an example of a Bell inequality in this setting with $m_\A=m_\B=2$.

Separable states satisfy all Bell inequalities with all measurements
by definition.
In a seminal paper~\cite{Wer-PRA89}, Werner disproved the converse:
there exists a quantum mixed state $\rho$ which is entangled but
satisfies all Bell inequalities.
To overcome the difficulty of proving these two properties of $\rho$, he
investigated states of very high symmetry now called Werner states.
Collins and Gisin~\cite{ColGis-JPA04} compared the strengths of Bell
inequalities by introducing a relevance relation between two Bell
inequalities, and they showed that a
Bell inequality named $\I_{3322}$ is relevant to the well-known CHSH
inequality.
Here relevance means that there is a quantum mixed state $\rho$ such
that $\rho$ satisfies the CHSH inequality (with all measurements) but
$\rho$ violates the $\I_{3322}$ inequality (with some measurements).
The state $\rho$ they found has less symmetry than the Werner states.

A test of relevance is a computationally difficult problem.
For one thing, to test relevance, one must tell whether a given state
satisfies a given Bell inequality for all measurements or not.
This can be cast as a bilinear semidefinite programming problem,
which is a hard optimization problem.
The ``see-saw iteration'' algorithm is used to solve it in
literature~\cite{WerWol-QIC01}.
Although it is not guaranteed to give the global optimum, multiple runs
with different initial solutions seem sufficient for many cases.
Another difficulty is to choose the appropriate state $\rho$.
Collins and Gisin overcome this difficulty by restricting states,
which we will describe in Section~\ref{sbsect:pre-relevance}.

Collins and Gisin showed numerically that the $\I_{3322}$ Bell inequality
is not relevant to the CHSH inequality for 2-level Werner states.
They posed an open problem~\cite{Gis-open19}: ``Find Bell inequalities
which are stronger than the CHSH inequalities in the sense that they
are violated by a wider range of Werner states.''
To answer this problem, we test 89 Bell inequalities for 2- and 3-level
isotropic states by using the see-saw iteration algorithm.
Isotropic states are a generalization of 2-level Werner states in that
they are convex combinations of a pure maximally entangled state and
the maximally mixed state.
The high symmetry of the isotropic states allows us to calculate the
maximum violation of the CHSH inequality by 3-level isotropic
states analytically.
The 89 inequalities used in the test are the Bell inequalities that
involve at most five measurements per party in the list of more than
200,000,000 tight Bell inequalities recently obtained by Avis, Imai,
Ito and Sasaki~\cite{AviImaItoSas:0404014,AviImaItoSas-JPA05} by using
a method known as triangular elimination.
We restrict computation to these 89 inequalities because the
optimization problem related to inequalities with many measurements is
difficult to solve.
As a result, we find five inequalities which are relevant to the
CHSH inequality for 3-level isotropic states.
They answer Collins and Gisin's problem where Werner states are
replaced by 3-level isotropic states.
We give empirical evidence that the five
inequalities are also relevant to the $\I_{3322}$ inequality.
To the best of our knowledge, no such Bell inequalities
were previously known.

The rest of the paper is organized as follows.
Section~\ref{sect:preliminaries} explains the necessary concepts.
Section~\ref{sect:inclusion} discusses inclusion relation, which is
used to prove irrelevance of a Bell inequality to another, and gives.
the inclusion relation among the Bell inequalities we used in our
experiments.
Section~\ref{sect:relevance} explains the method and the results of our
experiments to test relevance for 2- and 3-level isotropic states.
Section~\ref{sect:concluding} concludes the paper and mentions some
open problems.

\section{Preliminaries}  \label{sect:preliminaries}

\subsection{Bell inequalities} \label{sbsect:pre-bell}

We consider the following correlation experiment.
Suppose that two parties called Alice and Bob share a quantum state
$\rho$.
Alice has $m_\A$ choices
$\A_1,\dots,\A_{m_\A}$ of two-valued measurements and
Bob has $m_\B$ choices $\B_1,\dots,\B_{m_\B}$.
We call the two possible outcomes of the measurements 1 and 0.
The result of this correlation experiment can be represented by an
$(m_\A+m_\B+m_\A m_\B)$-dimensional vector $\vct{q}$, where for
$1\le i\le m_\A$ and $1\le j\le m_\B$, the variables
$q_{i0}$, $q_{0j}$ and $q_{ij}$ represent the probability
that the outcome of $\A_i$ is 1, that the outcome of $\B_j$ is 1, and
that two outcomes of both $\A_i$ and $\B_j$ are 1, respectively.

An inequality $\vct{a}^\trans\vct{q}\le a_0$, where $\vct{a}$ is an
$(m_\A+m_\B+m_\A m_\B)$-dimensional vector and $a_0$ is a scalar, is
called a \emph{Bell inequality} if it is satisfied for all separable
states $\rho$ and all choices of measurements
$\A_1,\dots,\A_{m_\A},\allowbreak\B_1,\dots,\B_{m_\B}$.
The nontrivial Bell inequality with the smallest values of $m_\A$ and
$m_\B$ is the CHSH inequality~\cite{ClaHorShiHol-PRL69}
\begin{equation}
  -q_{10}-q_{01}+q_{11}+q_{21}+q_{12}-q_{22}\le0 \label{eq:chsh}
\end{equation}
for $m_\A=m_\B=2$.

A Bell inequality is said to be \emph{tight} if it cannot
be written as a positive sum of two different Bell inequalities.
The CHSH inequality is an example of a tight Bell inequality.
Tight Bell inequalities are more useful as a test of the nonlocality
than the other Bell inequalities, since if a state violates a
non-tight Bell inequality $\vct{a}^\trans\vct{q}\le a_0$, then the
same state violates one of tight Bell inequalities which sum up to
$\vct{a}^\trans\vct{q}\le a_0$.

Throughout this paper, we denote a Bell inequality
$\vct{a}^\trans\vct{q}\le a_0$ by
\[
  \left(\begin{array}{cc||ccc}
               &         & (\A_1)  &\cdots& (\A_{m_\A})  \\
               &         &a_{10}   &\cdots&a_{m_\A0}     \\ \hline\hline
    (\B_1)     &a_{01}   &a_{11}   &\cdots&a_{m_\A1}     \\
     \vdots    & \vdots  & \vdots  &      & \vdots       \\
    (\B_{m_\B})&a_{0m_\B}&a_{1m_\B}&\cdots&a_{m_\A m_\B}
  \end{array} \right)\le a_0,
\]
following the notation by Collins and Gisin used in~\cite{ColGis-JPA04}
(with labels added to indicate which rows and columns correspond to
which measurements).
For example, the CHSH inequality~(\ref{eq:chsh}) is written as
\[
  \left(\begin{array}{cc||cc}
              &        & (\A_1) & (\A_2) \\
              &        &   -1   &    0   \\ \hline\hline
    (\B_1)    &   -1   &    1   &    1   \\
    (\B_2)    &    0   &    1   &   -1
  \end{array} \right)\le0.
\]

Another Bell inequality found by Pitowsky and
Svozil~\cite{PitSvo-PRA01} and named $\I_{3322}$ inequality by Collins
and Gisin~\cite{ColGis-JPA04} is written as
\begin{equation}
  \left(\begin{array}{cc||ccc}
              &        & (\A_1) & (\A_2) & (\A_3) \\
              &        &   -1   &    0   &    0   \\ \hline\hline
    (\B_1)    &   -2   &    1   &    1   &    1   \\
    (\B_2)    &   -1   &    1   &    1   &   -1   \\
    (\B_3)    &    0   &    1   &   -1   &    0
  \end{array} \right)\le0.
  \label{eq:i3322}
\end{equation}

Recently Avis, Imai, Ito and
Sasaki~\cite{AviImaItoSas:0404014,AviImaItoSas-JPA05} proposed a method
known as triangular elimination that can be used to generate tight
Bell inequalities from known tight inequalities for a well-studied related
polytope, known as the cut polytope.
They obtained a list of more than 200,000,000 tight Bell inequalities
by applying
triangular elimination to a list~\cite{SMAPO} of tight inequalities for the cut
polytope on 9 points, $\CutP_9$.
There are 89 Bell inequalities which involve five measurements per
party in the list, and they are used in this paper.
Among them are the CHSH inequality, the positive probability (trivial)
inequality, the $\I_{mm22}$ inequalities for $m=3,4,5$, the
$\I_{3422}^{(2)}$ inequality~\cite{ColGis-JPA04} and other unnamed
Bell inequalities.
We label the 89 inequalities as A1 to A89.
The list of these inequalities is available
online~\cite{ItoImaAvi:bell5-www}.

\subsection{Violation of a Bell inequality and bilinear semidefinite programming}
  \label{sbsect:pre-violation}

A test whether there exists a set of measurements violating a given
Bell inequality in a given state can be cast as a bilinear
semidefinite programming problem as follows.
Let $\rho$ be a density matrix in the $d\otimes d$ system and
$\vct{a}^\trans\vct{q}\le a_0$ be a Bell inequality.
Each measurement by Alice is represented by a \emph{positive operator
valued measure (POVM)} $(E_i,I-E_i)$, where $E_i$ is a Hermitian
$d\times d$ matrix such that both $E_i$ and $I-E_i$ are nonnegative
definite and $I$ is the identity matrix of size $d\times d$.
Similarly, each measurement by Bob is represented by a POVM
$(F_j,I-F_j)$.
For concise notation, we let $E_0=F_0=I$.
Then the test whether there exists a set of violating measurements or
not can be formulated as:
\begin{align}
  &\max\sum_{\substack{0\le i\le m_\A \\ 0\le j\le m_\B \\ (i,j)\ne(0,0)}}
   a_{ij}\tr(\rho(E_i\otimes F_j))-a_0
  \label{eq:optimization} \\
  &\text{where } E_0=F_0=I,\;\overline{E_i}^\trans=E_i,\;
   \overline{F_j}^\trans=F_j, \nonumber \\
  &O\preceq E_i{,}F_j\preceq I. \nonumber
\end{align}
Here the notation $X\preceq Y$ means that $Y-X$ is nonnegative
definite.
The optimal value of (\ref{eq:optimization}) is positive if and only
if there exist violating measurements, and if so, the optimal solution
gives the set of measurements that is maximally violating the given
Bell inequality in the given state.
If we fix one of the two groups of variables $\{E_1,\dots,E_{m_\A}\}$
and $\{F_1,\dots,F_{m_\B}\}$, (\ref{eq:optimization}) becomes a
semidefinite programming problem on the other group of variables.
In this respect, (\ref{eq:optimization}) can be seen as a variation of
bilinear programming~\cite{Kon-MP76} with semidefinite constraints.
The optimization problem (\ref{eq:optimization}) is NP-hard,
even for the case $d=1$, as follows from
results in ~\cite[Sections~5.1, 5.2]{DezLau:cut97}.

If $d=2$ and the inequality $\vct{a}^\trans\vct{q}\le a_0$ is the CHSH
inequality, then (\ref{eq:optimization}) can be solved
analytically~\cite{HorHorHor-PLA95}, hence the \emph{Horodecki
criterion}, a necessary and sufficient condition for a state $\rho$
in the $2\otimes2$ system to satisfy the CHSH inequality for all
measurements.
However, in general, the analytical solution of
(\ref{eq:optimization}) is not known.
This seems natural, given the difficulty of bilinear programming.
Section~2 of \cite{Kon-MP76} describes a hill-climbing algorithm which
computes a local optimum by fixing one of the two groups of variables
and solving the subproblem to optimize variables in the other groups
repeatedly, exchanging the role of the two groups in turn.
``See-saw iteration''~\cite{WerWol-QIC01} uses the same method
combined with the observation that in the case of
(\ref{eq:optimization}), each subproblem can be solved efficiently by
just computing the eigenvectors of a Hermitian $d\times d$ matrix.

There exists a set of \emph{projective} measurements
$E_1,\dots,E_{m_\A}$ and $F_1,\dots,F_{m_\B}$ which attains the
maximum of (\ref{eq:optimization}).
This fact is obtained from the proof of Theorem~5.4 in
\cite{CleHoyTonWat-CCC04} by Cleve, H{\o}yer, Toner and Watrous.
Though they prove the case where $\rho$ is also variable, the
relevant part in the proof is true even if the state is fixed.
See-saw iteration always produces projective measurements as a
candidate for the optimal measurements.

\subsection{Relevance relation} \label{sbsect:pre-relevance}

Collins and Gisin~\cite{ColGis-JPA04} introduced the notion of
relevance between two Bell inequalities and showed that the
Bell inequality~(\ref{eq:i3322}) named $\I_{3322}$ is relevant to the
well-known CHSH
inequality.
Here relevance means that there is a quantum mixed state $\rho$ such
that $\rho$ satisfies the CHSH inequality (with any measurements) but
$\rho$ violates the $\I_{3322}$ inequality (with some measurements).
They prove the relevance of the $\I_{3322}$ inequality to the CHSH
inequality by giving an explicit example of a state $\rho$ in the
$2\otimes2$ system which satisfies the CHSH inequality for all
measurements, and which violates the $\I_{3322}$
inequality for certain measurements.

Part of the difficulty of testing relevance comes from how to choose an
appropriate state $\rho$.
Even if we only consider the $2\otimes2$ system, the space of
mixed states is 15-dimensional.
Collins and Gisin overcome this difficulty by restricting the states
to those parameterized by two variables $\theta$ and $\alpha$:
$\rho(\theta,\alpha)
 =\alpha\lvert\varphi_\theta\rangle\langle\varphi_\theta\rvert
 +(1-\alpha)\lvert01\rangle\langle01\rvert$, where
$\lvert\varphi_\theta\rangle
 =\cos\theta\lvert00\rangle+\sin\theta\lvert11\rangle$.
For any $\theta$, the variable $\alpha$ can be maximized by using the
Horodecki criterion~\cite{HorHorHor-PLA95} to give a state
$\rho(\theta,\alpha_{\max})$ on the boundary of the set of the states
which satisfy the CHSH inequality for all measurements.
Then they compute the maximum violation of the $\I_{3322}$ inequality
by $\rho(\theta,\alpha_{\max})$
for various values of $\theta$, and find a state satisfying the
CHSH inequality but not the $\I_{3322}$ inequality.

\section{Inclusion relation} \label{sect:inclusion}

Before discussing relevance relations among Bell inequalities for
isotropic states, we need an introduction to inclusion relation among
these inequalities, which is used to distinguish ``obvious'' relevance
relations from the other relevance relations.

\subsection{Definition of inclusion relation}

Collins and Gisin~\cite{ColGis-JPA04} pointed out that the CHSH
inequality is irrelevant to the $\I_{3322}$ inequality
since if we pick the $\I_{3322}$ inequality and fix two measurements
$\A_3$ and $\B_1$ to the deterministic measurement whose result is
always 0, the inequality becomes the CHSH inequality.
Generalizing this argument, Avis, Imai, Ito and
Sasaki~\cite{AviImaItoSas-JPA05} introduced the notion of inclusion
relation between two Bell inequalities.
A Bell inequality $\vct{a}^\trans\vct{q}\le0$ \emph{includes}
another Bell inequality $\vct{b}^\trans\vct{q}\le0$ if we can obtain
the inequality $\vct{b}^\trans\vct{q}\le0$ by fixing some
measurements in the inequality $\vct{a}^\trans\vct{q}\le0$ to
deterministic ones (i.e.\ measurements whose result is always
1 or always 0).

Here we give a formal definition of the inclusion relation.
Let $\vct{a}^\trans\vct{q}\le0$ be a Bell inequality with
$m_A+m_B$ measurements and $\vct{b}^\trans\vct{q}\le0$ another
with $n_A+n_B$ measurements, and assume $m_A\ge n_A$ and $m_B\ge n_B$.
The inequality $\vct{a}^\trans\vct{q}\le0$ includes
$\vct{b}^\trans\vct{q}\le0$ if there exists a Bell inequality
$(\vct{a}')^\trans\vct{q}\le0$ equivalent to the inequality
$\vct{a}^\trans\vct{q}\le0$ such that
$a'_{ij}=b_{ij}$ for any $0\le i\le n_A$ and any $0\le j\le n_B$.
Here equivalence means that the inequality
$(\vct{a}')^\trans\vct{q}\le0$ can be obtained from another
$\vct{a}^\trans\vct{q}\le0$ by zero or more applications of party
exchange, observable exchange and value exchange.
See e.g.\ \cite{WerWol-PRA01} or \cite{ColGis-JPA04} for more about
equivalence of Bell inequalities.
Readers familiar with the cut polytope will recognize that inclusion
is a special case of collapsing~\cite[Section~26.4]{DezLau:cut97}.

By using this notion, a Bell inequality $\vct{a}^\trans\vct{q}\le0$ is
irrelevant to another Bell inequality $\vct{b}^\trans\vct{q}\le0$ if
the inequality $\vct{b}^\trans\vct{q}\le0$ includes the inequality
$\vct{a}^\trans\vct{q}\le0$.

\subsection{Inclusion relation between known Bell inequalities with at most 5 measurements per party} \label{sbsect:inclusion}

\begin{sidewaysfigure}
  \centering
  \includegraphics[width=\textwidth]{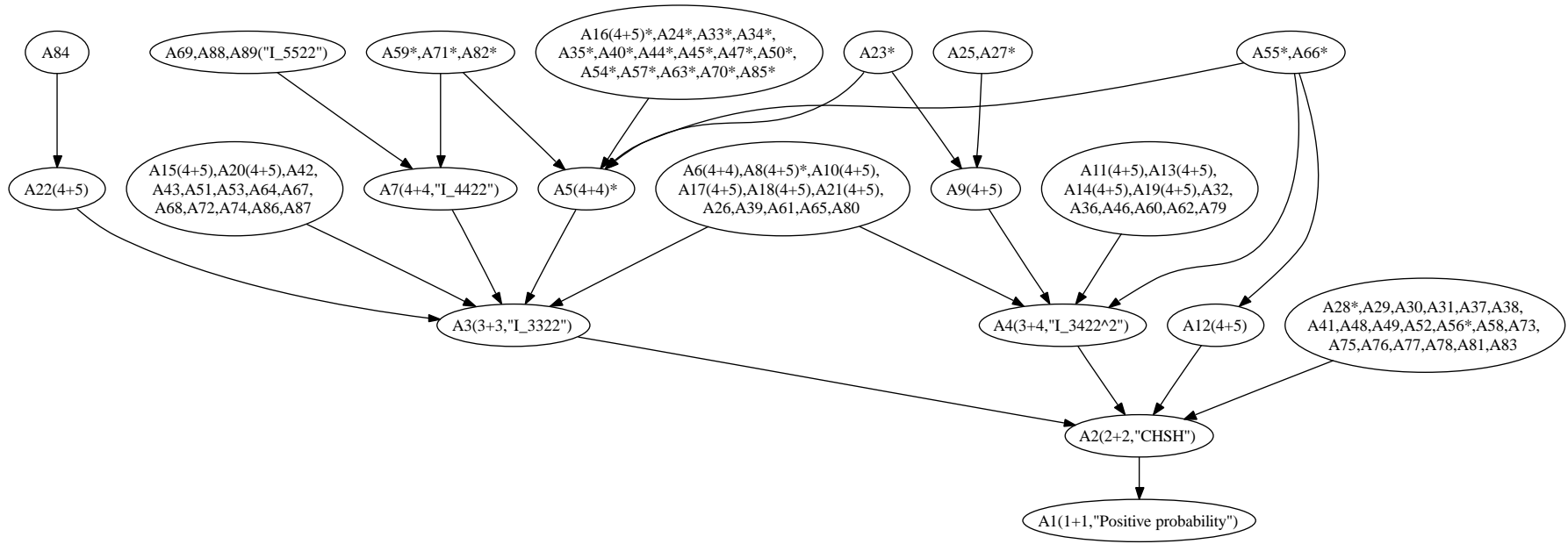}
  \caption{Inclusion relation among 89 Bell inequalities, with at most
    5 measurements per party, obtained by triangular elimination
    from facets of $\CutP_9$.
    An asterisk (*) on the right of the serial number indicates that
    the inequality is relevant to the CHSH inequality for $3\otimes3$
    isotropic states and that it is a candidate for being relevant to
    $\I_{3322}$.}
  \label{fig}
\end{sidewaysfigure}

We tested the inclusion relation among the 89 tight Bell inequalities
described in Section~\ref{sbsect:pre-bell}.
Figure~\ref{fig} on the last page shows the result.
In the figure, the serial number of each inequality is
shown with the number of measurements (omitted for inequalities with
$5+5$ measurements) and its name (if there is one).
An arc from one inequality to another means that the former includes
the latter.
Since the inclusion relation is transitive, the arcs which are derived by
other arcs are omitted.
An asterisk (*) on the right of the serial number indicates the
inequality is a candidate for being relevant to $\I_{3322}$.
Relevancy was tested empirically using the method described
in Section~\ref{ssect:comp}.

From the figure, one might be tempted to conjecture that the CHSH
inequality is included in all tight Bell inequalities other than the
positive probability inequality.
However, this is not true.
Enumeration of tight Bell inequalities with four measurements by each
party using the general convex hull computation package
lrs~\cite{Avi:lrs} takes an unrealistically long time, but in a partial
list, we have some counterexamples.
In the notation by Collins and Gisin, they are:
\begingroup
\allowdisplaybreaks
\let\hl\hline
\begin{align}
  &\left(\begin{array}{cc||cccc}
            &       & (\A_1) & (\A_2) & (\A_3) & (\A_4) \\
            &       &    0   &   -1   &   -1   &   -1   \\ \hl\hl
    (\B_1)  &  -1   &   -1   &    1   &    0   &    2   \\
    (\B_2)  &   0   &    0   &    1   &   -1   &   -1   \\
    (\B_3)  &  -1   &    1   &   -1   &    1   &    1   \\
    (\B_4)  &  -1   &   -1   &    1   &    2   &   -1
  \end{array} \right)\le0, \tag{$\I_{4422}^{(1)}$} \\
  &\left(\begin{array}{cc||cccc}
            &       & (\A_1) & (\A_2) & (\A_3) & (\A_4) \\
            &       &   -1   &    0   &   -1   &   -3   \\ \hl\hl
    (\B_1)  &   0   &    0   &    0   &   -1   &    1   \\
    (\B_2)  &  -1   &   -1   &    1   &    1   &    2   \\
    (\B_3)  &  -1   &    1   &   -1   &    2   &    1   \\
    (\B_4)  &   0   &    1   &   -1   &   -1   &    1
  \end{array} \right)\le0. \tag{$\I_{4422}^{(2)}$}
\end{align}
\endgroup

\section{Relevance for 2- and 3-level isotropic states}
\label{sect:relevance}

\subsection{Violation of a Bell inequality by isotropic states}

Let $\lvert\psi_d\rangle$ be a maximally entangled state in
$d\otimes d$ system:
\[
  \lvert\psi_d\rangle=\frac{1}{\sqrt{d}}
    (\lvert00\rangle+\lvert11\rangle+\dots+\lvert d-1,d-1\rangle).
\]
The \emph{$d$-level isotropic state~\cite{HorHorHor-PRL00} (or
$U\otimes U^*$-invariant state~\cite{HorHor-PRA99}) $\rho_d(\alpha)$
of parameter $0\le\alpha\le1$} is a state defined by:
\begin{align*}
  \rho_d(\alpha)
  &=\alpha\lvert\psi_d\rangle\langle\psi_d\rvert
   +(1-\alpha)\frac{I}{d^2} \\
  &=\frac{\alpha}{d}
    (\lvert00\rangle+\lvert11\rangle+\dots+\lvert d-1,d-1\rangle) \\
  &\hphantom{=\frac{\alpha}{d}}\;
    (\langle00\rvert+\langle11\rvert+\dots+\langle d-1,d-1\rvert)
   +\frac{1-\alpha}{d^2}I.
\end{align*}
With $\alpha=0$, $\rho_d(\alpha)$ is a maximally mixed state $I/d^2$,
which is separable and therefore satisfies all the Bell inequalities
for all measurements.
More generally, it is known that $\rho_d(\alpha)$ is separable if and
only if $\alpha\le1/(d+1)$~\cite{HorHor-PRA99}.
With $\alpha=1$, $\rho_d(\alpha)$ is a maximally entangled state
$\lvert\psi_d\rangle\langle\psi_d\rvert$.
Therefore $\rho_d(\alpha)$ represents a state in the middle between a
separable state and a maximally entangled state for general $\alpha$.

If two states $\rho$ and $\rho'$ satisfy a Bell inequality for all
measurements, then their convex combination $t\rho+(1-t)\rho'$ also
satisfies the same Bell inequality for all measurements.
This means that for any $d\ge2$ and any Bell inequality
$\vct{a}^\trans\vct{q}\le0$, there exists a real number
$0\le\alpha_{\max}\le1$ such that $\rho_d(\alpha)$ satisfies the
inequality $\vct{a}^\trans\vct{q}\le0$ for all measurements if and
only if $\alpha\le\alpha_{\max}$.
A smaller value of $\alpha_{\max}$ means that the Bell inequality is
more sensitive for isotropic states.

\subsection{Violation of the CHSH inequality by 3-level isotropic states}
  \label{sbsect:chsh}

In this section, we prove that the maximum violation of the CHSH
inequality by the 3-level isotropic state $\rho_3(\alpha)$ is given by
$\max\{0,\alpha(3\sqrt2+1)/9-4/9\}$.
As a corollary, the threshold $\alpha_{\max}$ for the CHSH inequality
with $d=3$ is equal to $\alpha_{\max}=4/(3\sqrt2+1)=0.76297427932$.

As we noted in Section~\ref{sbsect:pre-violation}, we can restrict
$E_1$, $E_2$, $F_1$ and $F_2$ to projective measurements in the
optimization problem~(\ref{eq:optimization}).
We consider the rank of measurements $E_1$, $E_2$, $F_1$ and $F_2$.
Since the CHSH inequality is not violated if any one of $E_1$, $E_2$,
$F_1$ and $F_2$ has rank zero or three, we only need to consider the
case where the four measurements $E_1$, $E_2$, $F_1$ and $F_2$ have
rank one or two.
Instead of considering all the combinations of ranks of the
measurements, we fix their rank to one and consider
the inequalities obtained by exchanging outcomes ``0'' and ``1'' of
some measurements in the CHSH inequality.
(In terms of the cut polytope, this transformation corresponds to
 switching~\cite[Section~26.3]{DezLau:cut97} of inequalities.
 See \cite{AviImaItoSas-JPA05} for details.)
For example, suppose that $E_1$ and $F_1$ have rank two and $E_2$ and
$F_2$ have rank one in the optimal set of measurements.
Then instead of the CHSH inequality in the form~(\ref{eq:chsh}), we
exchange the two outcomes of measurements $E_1$ and $F_1$ in the
inequality, and obtain (in the Collins-Gisin notation):
\begin{equation}
  \left(\begin{array}{cc||cc}
              &        & (\A_1) & (\A_2) \\
              &        &    0   &    1   \\ \hline\hline
    (\B_1)    &    0   &    1   &   -1   \\
    (\B_2)    &    1   &   -1   &   -1
  \end{array} \right)\le1, \label{eq:chsh-sw}
\end{equation}
with the four measurements of rank one.
We have $2^4=16$ possibilities for the ranks of the four measurements
and corresponding 16 inequalities transformed from (\ref{eq:chsh}).
These inequalities are identical to
either (\ref{eq:chsh}) or (\ref{eq:chsh-sw}) if it is relabelled
appropriately.
Therefore, we can assume the four measurements have rank one at the
expense of considering the inequality (\ref{eq:chsh-sw}) in addition
to (\ref{eq:chsh}).

We compute the maximum violation $V(\alpha)$ (resp.\ $V'(\alpha)$) of
the inequality (\ref{eq:chsh}) (resp.\ (\ref{eq:chsh-sw})) under the
assumption that the four measurements have rank one.
In the maximally mixed state $\rho_3(0)=I_9/9$, the violations of the
two inequalities are constant regardless of the actual measurements,
and they are:
\begin{align*}
  V(0)&=-q_{10}-q_{01}+q_{11}+q_{12}+q_{21}-q_{22} \\
  &=-1/3-1/3+1/9+1/9+1/9-1/9=-4/9, \\
  V'(0)&=q_{20}+q_{02}+q_{11}-q_{12}-q_{21}-q_{22}-1 \\
  &=1/3+1/3+1/9-1/9-1/9-1/9-1=-5/9.
\end{align*}
Since the violations of the inequalities are constant in the state
$\rho_3(0)$, the maximum violation in the state $\rho_3(\alpha)$ is
achieved by the optimal set of measurements in the state $\rho_3(1)$,
$V(\alpha)=\alpha V(1)+(1-\alpha)V(0)$ and
$V'(\alpha)=\alpha V'(1)+(1-\alpha)V'(0)$.
Therefore, what remains is to compute the values of $V(1)$ and
$V'(1)$.

To obtain the value of $V(1)$, let
$E_i=\lvert\varphi_{1i}\rangle\langle\varphi_{1i}\rvert$,
$F_j=\lvert\varphi_{2j}\rangle\langle\varphi_{2j}\rvert$,
$\lvert\varphi_{1i}\rangle=x_{i0}\lvert0\rangle+x_{i1}\lvert1\rangle
+x_{i2}\lvert2\rangle$ and
$\lvert\varphi_{2j}\rangle=\overline{y_{j0}}\lvert0\rangle
+\overline{y_{j1}}\lvert1\rangle+\overline{y_{j2}}\lvert2\rangle$.
Note that $\vct{x}_1$, $\vct{x}_2$, $\vct{y}_1$ and $\vct{y}_2$ are
unit vectors in $\CC^3$.
Using them, the violations of the inequality (\ref{eq:chsh}) is equal
to
\begin{equation}
  -\frac23
  +\frac13( \lvert\vct{x}_1\cdot\vct{y}_1\rvert^2
           +\lvert\vct{x}_1\cdot\vct{y}_2\rvert^2
           +\lvert\vct{x}_2\cdot\vct{y}_1\rvert^2
           -\lvert\vct{x}_2\cdot\vct{y}_2\rvert^2),
  \label{eq:chsh-violation}
\end{equation}
If we fix $\vct{y}_1$ and $\vct{y}_2$ arbitrarily, then optimization
of $\vct{x}_1$ and $\vct{x}_2$ in (\ref{eq:chsh-violation}) can be
performed separately.
Since (\ref{eq:chsh-violation}) depends only on the inner products of
the vectors and not the vectors themselves, we can replace the vectors
$\vct{x}_1$ and $\vct{x}_2$ with their projection onto the subspace
spanned by $\vct{y}_1$ and $\vct{y}_2$.
This means that we can consider the four vectors $\vct{x}_1$,
$\vct{x}_2$, $\vct{y}_1$ and $\vct{y}_2$ are vectors in $\CC^2$ whose
lengths are at most one.
Then the Tsirelson inequality~\cite{Tsi-JSovM87,CheBar-JPA96} tells the
maximum of
$ \lvert\vct{x}_1\cdot\vct{y}_1\rvert^2
 +\lvert\vct{x}_1\cdot\vct{y}_2\rvert^2
 +\lvert\vct{x}_2\cdot\vct{y}_1\rvert^2
 -\lvert\vct{x}_2\cdot\vct{y}_2\rvert^2$
is equal to $\sqrt{2}+1$, and the vectors giving this maximum are
$\lvert\varphi_{11}\rangle=\cos(\pi/4)\lvert0\rangle+\sin(\pi/4)\lvert1\rangle$,
$\lvert\varphi_{12}\rangle=\lvert0\rangle$,
$\lvert\varphi_{21}\rangle=\cos(\pi/8)\lvert0\rangle+\sin(\pi/8)\lvert1\rangle$
and
$\lvert\varphi_{22}\rangle=\cos(3\pi/8)\lvert0\rangle+\sin(3\pi/8)\lvert1\rangle$.
The violation of (\ref{eq:chsh}) is $V(1)=(\sqrt2-1)/3=0.138071$, and
$V(\alpha)=(1-\alpha)(-4/9)+\alpha(\sqrt2-1)/3=\alpha(3\sqrt2+1)/9-4/9$.

By a similar argument, we can compute the value of $V'(1)$.
Using the same definition for $\vct{x}_1$, $\vct{x}_2$, $\vct{y}_1$
and $\vct{y}_2$, the violation of the inequality (\ref{eq:chsh-sw}) is
given by
\begin{equation}
  -\frac43
  +\frac13( \lvert\vct{x}_1\cdot\vct{y}_1\rvert^2
           -\lvert\vct{x}_1\cdot\vct{y}_2\rvert^2
           -\lvert\vct{x}_2\cdot\vct{y}_1\rvert^2
           -\lvert\vct{x}_2\cdot\vct{y}_2\rvert^2).
  \label{eq:chsh-violation-sw}
\end{equation}
The maximum of (\ref{eq:chsh-violation-sw}) is equal to $-1$, and it
is achieved by setting
$\lvert\varphi_{11}\rangle=\lvert\varphi_{21}\rangle=\lvert0\rangle$,
$\lvert\varphi_{12}\rangle=\lvert1\rangle$ and
$\lvert\varphi_{22}\rangle=\lvert2\rangle$.
Therefore $V'(1)=-1$ and $V'(\alpha)=-14\alpha/9-5/9<0$.
This means the inequality (\ref{eq:chsh-sw}) is never violated under
the assumption that the four measurements have rank one.

Removing the assumption of the ranks of the measurements, we obtain
that the maximum violation of the CHSH inequality in the state
$\rho_3(\alpha)$ is given by
$\max\{0,V(\alpha),V'(\alpha)\}=\max\{0,\alpha(3\sqrt2+1)/9-4/9\}$.

\subsection{Computation of violation of Bell inequalities
  with at most 5 measurements per party} \label{ssect:comp}

We performed preliminary experiments to compute an upper bound on the
value of $\alpha_{\max}$ with $d=2$ and $d=3$ for the 89 inequalities
described in Section~\ref{sbsect:pre-bell}.
The see-saw iteration algorithm finds a candidate for the optimal
solution of (\ref{eq:optimization}).
When $0\le\alpha\le1$ is given, we can use this search algorithm to
tell whether $\alpha_{\max}<\alpha$ (if violating measurements are
found) or $\alpha_{\max}\ge\alpha$ (otherwise), if we ignore the
possibility that the hill-climbing search fails to find the global
optimum.
This allows us to compute the value of $\alpha_{\max}$ by binary
search.
In reality, the hill-climbing search sometimes fails to find the
global optimum, and if it finds violating measurements then it surely
means $\alpha_{\max}<\alpha$, whereas if it does not find violating
measurements then it does not necessarily mean
$\alpha_{\max}\ge\alpha$.
Therefore, the value given by binary search is not necessarily the
true value of $\alpha_{\max}$ but an upper bound on it.

In each step of the binary search, we performed a see-saw iteration with
1,000 random initial measurements and picked the solution giving the
maximum in the 1,000 trials.
To compute eigenvalues and eigenvectors of $3\times3$ Hermitian
matrix, we used LAPACK~\cite{lapack} with
ATLAS~\cite{WhaPetDon-PC01,atlas}.
All computations were performed using double-precision floating
arithmetic.
Due to numerical error, the computation indicates a small positive
violation even if the state does not violate the inequality.
Therefore, we only consider violation greater than $10^{-13}$
significant.

For $d=2$, the computation gave an upper bound $0.70711$ for all
inequalities except for the positive probability inequality.
(For the positive probability inequality we have $\alpha_{\max}=1$
since it is satisfied by any quantum state.)
It is known that in the case $d=2$, the CHSH inequality is satisfied if and only
if $\alpha\le1/\sqrt2=0.70711$ from the Horodecki
criterion~\cite{HorHorHor-PLA95}.
These results suggest that there may not be any Bell inequalities
relevant to the CHSH inequality for 2-level isotropic states,
indicating the negative answer to
Gisin's problem~\cite{Gis-open19} in the case of 2-level system.

\begin{table*}
  \centering
  \caption{Upper bound of the value of $\alpha_{\max}$ obtained by the
    experiments.}
  \label{tab}
  \begin{tabular}{|l|cl|cl|} \hline
    \multicolumn{1}{|c|}{$\alpha_{\max}$} &
      \multicolumn{2}{|c|}{Bell inequality} &
      \multicolumn{2}{|c|}{Original cut polytope inequality} \\\hline
    0.7447198434 & A28 &                      &     7 & \\
    0.7453308276 & A27 &                      &     6 & \\
    0.7553800191 & A5  &                      &     8 & (Par(7), parachute ineq.) \\
    0.7557816805 & A56 &                      &    89 & \\
    0.7614396336 & A8  &                      &     2 & (Pentagonal ineq.) \\
    0.7629742793 & A3  &($\I_{3322}$)         &     2 & (Pentagonal ineq.) \\
    0.7629742793 & A2  &(CHSH)                &     1 & (Triangle ineq.) \\
    1            & A1  &(Positive probability)&     1 & (Triangle ineq.) \\\hline
  \end{tabular}
\end{table*}

We performed the same computation for $d=3$.
This time some Bell inequalities gave a smaller value of
$\alpha_{\max}$ than the CHSH inequality did.
Some of them gave a small value of $\alpha_{\max}$ simply because it
includes another such inequality.
Filtering them out, we identified five inequalities which are candidates
for being relevant to the CHSH inequality for the 3-level isotropic
states.
Rows and columns in bold font indicate that they correspond to
nodes added by triangular elimination.

\begingroup
\allowdisplaybreaks
\let\hl\hline
\def\b#1{\bm{(#1)}}
\begin{alignat*}{2}
  &\text{A28:} & \quad
  &\left(\begin{array}{cc||ccccc}
            &       & (\A_1) & (\A_2) & (\A_3) & (\A_4) &\b{\A_5}\\
            &       &   -2   &   -1   &   -1   &    0   &\bm{ 0} \\ \hl\hl
    (\B_1)  &  -2   &    1   &    0   &    1   &    1   &\bm{ 1} \\
    (\B_2)  &  -1   &    0   &    1   &    1   &    1   &\bm{-1} \\
    (\B_3)  &  -1   &    1   &    1   &   -1   &    0   &\bm{ 0} \\
    (\B_4)  &   0   &    1   &    1   &    0   &   -1   &\bm{ 0} \\
    \b{\B_5}&\bm{ 0}&\bm{ 1} &\bm{-1} &\bm{ 0} &\bm{ 0} &\bm{ 0}
  \end{array} \right)\le0, \\
  &\text{A27:} & \quad
  &\left(\begin{array}{cc||ccccc}
            &       & (\A_1) & (\A_2) & (\A_3) & (\A_4) &\b{\A_5}\\
            &       &   -1   &    0   &    0   &   -1   &\bm{-1} \\ \hl\hl
    (\B_1)  &  -2   &    1   &    1   &    1   &    0   &\bm{ 0} \\
    (\B_2)  &   0   &    1   &    0   &   -1   &   -1   &\bm{ 1} \\
    (\B_3)  &  -1   &    0   &   -1   &    1   &    1   &\bm{ 1} \\
    (\B_4)  &  -1   &   -1   &    1   &    0   &    1   &\bm{ 0} \\
    \b{\B_5}&\bm{-1}&\bm{ 1} &\bm{ 0} &\bm{ 0} &\bm{ 1} &\bm{ 0}
  \end{array} \right)\le0, \\
  &\text{A5:} & \quad
  &\left(\begin{array}{cc||cccc}
            &       & (\A_1) & (\A_2) & (\A_3) &\b{\A_4}\\
            &       &    0   &    0   &   -1   &\bm{-1} \\ \hl\hl
    (\B_1)  &  -2   &    1   &    1   &    1   &\bm{ 0} \\
    (\B_2)  &  -1   &    1   &   -1   &    0   &\bm{ 1} \\
    (\B_3)  &  -1   &   -1   &    1   &    1   &\bm{ 1} \\
    \b{\B_4}&\bm{ 0}&\bm{ 0} &\bm{-1} &\bm{ 1} &\bm{ 0}
  \end{array} \right)\le0, \\
  &\text{A56:} & \quad
  &\left(\begin{array}{cc||ccccc}
            &       & (\A_1) & (\A_2) & (\A_3) & (\A_4) &\b{\A_5}\\
            &       &   -1   &    0   &    0   &   -2   &\bm{-2} \\ \hl\hl
    (\B_1)  &  -1   &    0   &    1   &   -1   &    1   &\bm{ 0} \\
    (\B_2)  &   0   &    1   &    0   &   -1   &    1   &\bm{ 0} \\
    (\B_3)  &   0   &   -1   &   -1   &   -1   &    1   &\bm{ 2} \\
    (\B_4)  &  -2   &    1   &    1   &    1   &   -1   &\bm{ 2} \\
    \b{\B_5}&\bm{-2}&\bm{ 0} &\bm{ 0} &\bm{ 2} &\bm{ 2} &\bm{ 0}
  \end{array} \right)\le0, \\
  &\text{A8:} & \quad
  &\left(\begin{array}{cc||cccc}
            &       & (\A_1) & (\A_2) & (\A_3) &\b{\A_4}\\
            &       &    0   &   -1   &   -2   &\bm{ 0} \\ \hl\hl
    (\B_1)  &  -1   &    1   &    1   &    1   &\bm{-1} \\
    (\B_2)  &  -2   &    1   &    1   &    1   &\bm{ 1} \\
    \b{\B_3}&\bm{ 0}&\bm{-1} &\bm{ 1} &\bm{ 0} &\bm{ 0} \\
    \b{\B_4}&\bm{ 0}&\bm{-1} &\bm{ 0} &\bm{ 1} &\bm{ 0} \\
    \b{\B_5}&\bm{ 0}&\bm{ 0} &\bm{-1} &\bm{ 1} &\bm{ 0}
  \end{array} \right)\le0.
\end{alignat*}
\endgroup

Adding the CHSH and the $\I_{3322}$ inequalities, we performed the
experiments with 50,000 initial solutions with the seven inequalities.
Table~\ref{tab} summarizes the results we obtained.
In Table~\ref{tab}, the column labeled ``Original cut polytope inequality'' shows
the facet inequality of $\CutP_9$ to which triangular elimination is
applied.
The number corresponds to the serial number of the facet in
\texttt{cut9.gz} of \cite{SMAPO}.
For the CHSH inequality, the obtained upper bound $0.76298$ is
consistent with the theoretical value $4/(3\sqrt2+1)=0.762974$ proved
in Section~\ref{sbsect:chsh}.
The $\I_{3322}$ inequality gave the same upper bound as the CHSH
inequality.
Besides, in the optimal measurements with $\alpha$ near
$4/(3\sqrt2+1)$, the matrices $E_3$ and $F_1$ are zero, corresponding
to the fact that the $\I_{3322}$ inequality includes the CHSH
inequality.
This is consistent with Collins and Gisin's
observation~\cite{ColGis-JPA04} in the $2\otimes2$ system that the
$\I_{3322}$ inequality is not better than the CHSH inequality for
states with high symmetry.

Five Bell inequalities A28, A27, A5, A56 and A8 gave a smaller value of
$\alpha_{\max}$ than $4/(3\sqrt2+1)$.
The set of measurements giving optimal violation for these Bell
inequalities with $\alpha$ slightly larger than the computed
value of $\alpha_{\max}$ is given in the Appendix.

These Bell inequalities are relevant to the CHSH inequality.
As a result, Bell inequalities including any of them are also relevant
to the CHSH inequality.
Moreover, if the true value of $\alpha_{\max}$ for the $\I_{3322}$
inequality is $4/(3\sqrt2+1)$, then these five Bell inequalities are
also relevant to the $\I_{3322}$ inequality.
We make the following conjecture.

\begin{cnjc}
  The state $\rho_3(4/(3\sqrt2+1))$ satisfies the $\I_{3322}$
  inequality for all measurements.
  In other words, $\alpha_{\max}=4/(3\sqrt2+1)$ for the $\I_{3322}$
  inequality in the case of $d=3$.
\end{cnjc}

To support this conjecture, we searched for the optimal measurements
for the $\I_{3322}$ inequality in the states $\rho_3(\alpha)$ with
$\alpha=\alpha_+=0.7629742794>4/(3\sqrt2+1)$ and
$\alpha=\alpha_-=0.7629742793<4/(3\sqrt2+1)$, using see-saw iteration
algorithm with random initial solutions.
With $\alpha=\alpha_+$, 100 out of 633 trials gave a violation greater
than $10^{-13}$, whereas with $\alpha=\alpha_-$, none of 50,000 trials
gave a violation greater than $3{\times}10^{-15}$.
Considering numerical error in computation, we consider that this
result can be seen as an evidence that the $\I_{3322}$ inequality
behaves differently in the state $\rho_3(\alpha)$ depending on whether
$\alpha$ is greater or less than $4/(3\sqrt2+1)$.

\section{Concluding remarks} \label{sect:concluding}

We used numerical optimization to show that certain 
Bell inequalities are relevant to
the CHSH inequality for isotropic states.
No Bell inequalities relevant to the CHSH inequality were found for
2-level isotropic states.
This supports Collins and Gisin's conjecture in \cite{ColGis-JPA04}
that no such Bell inequalities exist.
For 3-level isotropic states, however, five Bell inequalities relevant
to the CHSH inequality were found.
The results of numerical experiments were given to support the
conjecture that they are also relevant for the $\I_{3322}$ inequality.

The violation of the CHSH inequality by 3-level isotropic states was
shown by using Tsirelson's inequality.
Cleve, H\o yer, Toner and Watrous~\cite{CleHoyTonWat-CCC04}
generalize Tsirelson's inequality to Bell inequalities corresponding
to ``XOR games,'' which do not depend on individual variables
$q_{i0},q_{0j},q_{ij}$ but only involves combinations in the form
$x_{ij}=q_{i0}+q_{0j}-2q_{ij}$.
Unfortunately, the $\I_{3322}$ inequality is not such an inequality,
and we cannot use the result there to prove the theoretical
value of $\alpha_{\max}$ for the $\I_{3322}$ inequality.
Among the five Bell inequalities relevant to the CHSH inequality for
3-level isotropic states, the inequality A8,
which can be written as
$-\sum_{i=1,2}\sum_{j=1,2,3}x_{ij}+x_{13}-x_{23}+x_{14}-x_{34}
 +x_{25}-x_{35}+x_{41}-x_{42}\le0$,
is the only one that corresponds to an XOR game.
An important open problem is to generalize Cleve, H\o yer, Toner and
Watrous's result to cover Bell inequalities which do not
correspond to XOR games.

\begin{acknowledgments}
  The first author is supported by the Grant-in-Aid for JSPS Fellows.
\end{acknowledgments}

\appendix
\begin{widetext} \footnotesize
\section{Optimal measurements computed for each inequalities}

\noindent A28:
\begin{align*}
  E_1&=I-\lvert\varphi_{11}\rangle\langle\varphi_{11}\rvert, &\quad
  &\lvert\varphi_{11}\rangle=0.819512\lvert0\rangle+(- 0.181891 - 0.067213i)\lvert1\rangle+(0.239561 + 0.483124i)\lvert2\rangle, \\
  E_2&=\lvert\varphi_{12}\rangle\langle\varphi_{12}\rvert, &\quad
  &\lvert\varphi_{12}\rangle=0.391928\lvert0\rangle+(0.546808 - 0.330668i)\lvert1\rangle+(- 0.064601 + 0.658695i)\lvert2\rangle, \\
  E_3&=\lvert\varphi_{13}\rangle\langle\varphi_{13}\rvert, &\quad
  &\lvert\varphi_{13}\rangle=0.585206\lvert0\rangle+(0.266618 - 0.150612i)\lvert1\rangle+(0.721307 - 0.208519i)\lvert2\rangle, \\
  E_4&=\lvert\varphi_{14}\rangle\langle\varphi_{14}\rvert, &\quad
  &\lvert\varphi_{14}\rangle=0.696701\lvert0\rangle+(0.109760 + 0.562926i)\lvert1\rangle+(0.269399 - 0.336302i)\lvert2\rangle, \\
  E_5&=I-\lvert\varphi_{15}\rangle\langle\varphi_{15}\rvert, &\quad
  &\lvert\varphi_{15}\rangle=0.745551\lvert0\rangle+(0.060720 - 0.038486i)\lvert1\rangle+(0.610743 + 0.256863i)\lvert2\rangle, \\
  F_1&=I-\lvert\varphi_{21}\rangle\langle\varphi_{21}\rvert, &\quad
  &\lvert\varphi_{21}\rangle=0.665942\lvert0\rangle+(0.124951 + 0.288249i)\lvert1\rangle+(0.306094 - 0.603430i)\lvert2\rangle, \\
  F_2&=I-\lvert\varphi_{22}\rangle\langle\varphi_{22}\rvert, &\quad
  &\lvert\varphi_{22}\rangle=0.794583\lvert0\rangle+(- 0.503910 - 0.071325i)\lvert1\rangle+(- 0.075809 - 0.322300i)\lvert2\rangle, \\
  F_3&=I-\lvert\varphi_{23}\rangle\langle\varphi_{23}\rvert, &\quad
  &\lvert\varphi_{23}\rangle=0.738612\lvert0\rangle+(0.143632 - 0.211840i)\lvert1\rangle+(0.594179 + 0.189467i)\lvert2\rangle, \\
  F_4&=\lvert\varphi_{24}\rangle\langle\varphi_{24}\rvert, &\quad
  &\lvert\varphi_{24}\rangle=0.314299\lvert0\rangle+(0.087381 + 0.592536i)\lvert1\rangle+(0.427166 + 0.600009i)\lvert2\rangle, \\
  F_5&=I-\lvert\varphi_{25}\rangle\langle\varphi_{25}\rvert, &\quad
  &\lvert\varphi_{25}\rangle=0.745551\lvert0\rangle+(0.060720 + 0.038486i)\lvert1\rangle+(0.610743 - 0.256863i)\lvert2\rangle
\end{align*}

\noindent A27:
\begin{align*}
  E_1&=\lvert\varphi_{11}\rangle\langle\varphi_{11}\rvert, &\quad
  &\lvert\varphi_{11}\rangle=0.512740\lvert0\rangle+(0.141298 - 0.367921i)\lvert1\rangle+(0.118341 - 0.753500i)\lvert2\rangle, \\
  E_2&=I-\lvert\varphi_{12}\rangle\langle\varphi_{12}\rvert, &\quad
  &\lvert\varphi_{12}\rangle=0.429346\lvert0\rangle+(0.490358 + 0.190555i)\lvert1\rangle+(- 0.588595 - 0.438697i)\lvert2\rangle, \\
  E_3&=I-\lvert\varphi_{13}\rangle\langle\varphi_{13}\rvert, &\quad
  &\lvert\varphi_{13}\rangle=0.649098\lvert0\rangle+(- 0.034498 + 0.390106i)\lvert1\rangle+(0.648622 + 0.067734i)\lvert2\rangle, \\
  E_4&=\lvert\varphi_{14}\rangle\langle\varphi_{14}\rvert, &\quad
  &\lvert\varphi_{14}\rangle=0.782874\lvert0\rangle+(- 0.199336 - 0.104823i)\lvert1\rangle+(- 0.579621 + 0.020651i)\lvert2\rangle, \\
  E_5&=\lvert\varphi_{15}\rangle\langle\varphi_{15}\rvert, &\quad
  &\lvert\varphi_{15}\rangle=0.504711\lvert0\rangle+(0.266955 - 0.029362i)\lvert1\rangle+(- 0.176172 - 0.801313i)\lvert2\rangle, \\
  F_1&=\lvert\varphi_{21}\rangle\langle\varphi_{21}\rvert, &\quad
  &\lvert\varphi_{21}\rangle=0.477430\lvert0\rangle+(- 0.243408 + 0.631106i)\lvert1\rangle+(- 0.024181 + 0.560297i)\lvert2\rangle, \\
  F_2&=\lvert\varphi_{22}\rangle\langle\varphi_{22}\rvert, &\quad
  &\lvert\varphi_{22}\rangle=0.521997\lvert0\rangle+(0.270933 - 0.132987i)\lvert1\rangle+(0.586914 + 0.540334i)\lvert2\rangle, \\
  F_3&=\lvert\varphi_{23}\rangle\langle\varphi_{23}\rvert, &\quad
  &\lvert\varphi_{23}\rangle=0.631718\lvert0\rangle+(0.176373 + 0.079451i)\lvert1\rangle+(- 0.678537 + 0.321093i)\lvert2\rangle, \\
  F_4&=\lvert\varphi_{24}\rangle\langle\varphi_{24}\rvert, &\quad
  &\lvert\varphi_{24}\rangle=0.839814\lvert0\rangle+(- 0.361305 - 0.101706i)\lvert1\rangle+(- 0.207777 - 0.332648i)\lvert2\rangle, \\
  F_5&=\lvert\varphi_{25}\rangle\langle\varphi_{25}\rvert, &\quad
  &\lvert\varphi_{25}\rangle=0.634648\lvert0\rangle+(- 0.135288 + 0.308277i)\lvert1\rangle+(- 0.492423 + 0.491328i)\lvert2\rangle
\end{align*}

\noindent A5:
\begin{align*}
  E_1&=I-\lvert\varphi_{11}\rangle\langle\varphi_{11}\rvert, &\quad
  &\lvert\varphi_{11}\rangle=0.079911\lvert0\rangle+(0.347597 - 0.352563i)\lvert1\rangle+(0.852394 + 0.148034i)\lvert2\rangle, \\
  E_2&=\lvert\varphi_{12}\rangle\langle\varphi_{12}\rvert, &\quad
  &\lvert\varphi_{12}\rangle=0.466812\lvert0\rangle+(0.336458 - 0.338316i)\lvert1\rangle+(0.063365 - 0.741896i)\lvert2\rangle, \\
  E_3&=I-\lvert\varphi_{13}\rangle\langle\varphi_{13}\rvert, &\quad
  &\lvert\varphi_{13}\rangle=0.700997\lvert0\rangle+(- 0.090375 + 0.325520i)\lvert1\rangle+(0.625759 - 0.053829i)\lvert2\rangle, \\
  E_4&=\lvert\varphi_{14}\rangle\langle\varphi_{14}\rvert, &\quad
  &\lvert\varphi_{14}\rangle=0.569742\lvert0\rangle+(- 0.703808 - 0.061209i)\lvert1\rangle+(- 0.405767 - 0.107957i)\lvert2\rangle, \\
  F_1&=\lvert\varphi_{21}\rangle\langle\varphi_{21}\rvert, &\quad
  &\lvert\varphi_{21}\rangle=0.611974\lvert0\rangle+(0.261472 + 0.553836i)\lvert1\rangle+(- 0.402289 + 0.297574i)\lvert2\rangle, \\
  F_2&=\lvert\varphi_{22}\rangle\langle\varphi_{22}\rvert, &\quad
  &\lvert\varphi_{22}\rangle=0.743739\lvert0\rangle+(- 0.644052 - 0.121119i)\lvert1\rangle+(- 0.050055 - 0.121959i)\lvert2\rangle, \\
  F_3&=\lvert\varphi_{23}\rangle\langle\varphi_{23}\rvert, &\quad
  &\lvert\varphi_{23}\rangle=0.327181\lvert0\rangle+(- 0.492820 + 0.363796i)\lvert1\rangle+(- 0.442899 + 0.567075i)\lvert2\rangle, \\
  F_4&=I-\lvert\varphi_{24}\rangle\langle\varphi_{24}\rvert, &\quad
  &\lvert\varphi_{24}\rangle=0.558366\lvert0\rangle+(0.295353 - 0.157594i)\lvert1\rangle+(0.593099 + 0.473699i)\lvert2\rangle
\end{align*}

\noindent A56:
\begin{align*}
  E_1&=\lvert\varphi_{11}\rangle\langle\varphi_{11}\rvert, &\quad
  &\lvert\varphi_{11}\rangle=0.764669\lvert0\rangle+(0.520735 - 0.023147i)\lvert1\rangle+(0.314448 - 0.211429i)\lvert2\rangle, \\
  E_2&=I-\lvert\varphi_{12}\rangle\langle\varphi_{12}\rvert, &\quad
  &\lvert\varphi_{12}\rangle=0.523087\lvert0\rangle+(- 0.660068 + 0.130414i)\lvert1\rangle+(0.115043 + 0.510340i)\lvert2\rangle, \\
  E_3&=I-\lvert\varphi_{13}\rangle\langle\varphi_{13}\rvert, &\quad
  &\lvert\varphi_{13}\rangle=0.651881\lvert0\rangle+(0.010176 - 0.025750i)\lvert1\rangle+(- 0.599260 + 0.463866i)\lvert2\rangle, \\
  E_4&=I-\lvert\varphi_{14}\rangle\langle\varphi_{14}\rvert, &\quad
  &\lvert\varphi_{14}\rangle=0.480244\lvert0\rangle+(0.435821 - 0.476742i)\lvert1\rangle+(0.370530 + 0.463520i)\lvert2\rangle, \\
  E_5&=I-\lvert\varphi_{15}\rangle\langle\varphi_{15}\rvert, &\quad
  &\lvert\varphi_{15}\rangle=0.484893\lvert0\rangle+(0.214118 + 0.403736i)\lvert1\rangle+(0.401826 + 0.628144i)\lvert2\rangle, \\
  F_1&=\lvert\varphi_{21}\rangle\langle\varphi_{21}\rvert, &\quad
  &\lvert\varphi_{21}\rangle=0.704822\lvert0\rangle+(0.050276 - 0.044858i)\lvert1\rangle+(- 0.676460 + 0.202702i)\lvert2\rangle, \\
  F_2&=I-\lvert\varphi_{22}\rangle\langle\varphi_{22}\rvert, &\quad
  &\lvert\varphi_{22}\rangle=0.279921\lvert0\rangle+(- 0.406294 + 0.685472i)\lvert1\rangle+(0.534341 + 0.034308i)\lvert2\rangle, \\
  F_3&=I-\lvert\varphi_{23}\rangle\langle\varphi_{23}\rvert, &\quad
  &\lvert\varphi_{23}\rangle=0.580814\lvert0\rangle+(0.563163 + 0.064963i)\lvert1\rangle+(0.561359 - 0.161735i)\lvert2\rangle, \\
  F_4&=I-\lvert\varphi_{24}\rangle\langle\varphi_{24}\rvert, &\quad
  &\lvert\varphi_{24}\rangle=0.522791\lvert0\rangle+(- 0.366663 - 0.240466i)\lvert1\rangle+(- 0.161766 - 0.712921i)\lvert2\rangle, \\
  F_5&=I-\lvert\varphi_{25}\rangle\langle\varphi_{25}\rvert, &\quad
  &\lvert\varphi_{25}\rangle=0.575083\lvert0\rangle+(0.352241 + 0.118045i)\lvert1\rangle+(- 0.170766 - 0.708598i)\lvert2\rangle
\end{align*}

\noindent A8:
\begin{align*}
  E_1&=\lvert\varphi_{11}\rangle\langle\varphi_{11}\rvert, &\quad
  &\lvert\varphi_{11}\rangle=0.589845\lvert0\rangle+(0.252414 - 0.592962i)\lvert1\rangle+(- 0.067286 + 0.481911i)\lvert2\rangle, \\
  E_2&=\lvert\varphi_{12}\rangle\langle\varphi_{12}\rvert, &\quad
  &\lvert\varphi_{12}\rangle=0.571429\lvert0\rangle+(- 0.328221 - 0.214531i)\lvert1\rangle+(0.352103 + 0.629079i)\lvert2\rangle, \\
  E_3&=I-\lvert\varphi_{13}\rangle\langle\varphi_{13}\rvert, &\quad
  &\lvert\varphi_{13}\rangle=0.789596\lvert0\rangle+(0.397845 + 0.124284i)\lvert1\rangle+(0.373987 + 0.250887i)\lvert2\rangle, \\
  E_4&=I-\lvert\varphi_{14}\rangle\langle\varphi_{14}\rvert, &\quad
  &\lvert\varphi_{14}\rangle=0.588353\lvert0\rangle+(- 0.068306 - 0.217513i)\lvert1\rangle+(- 0.748446 + 0.204184i)\lvert2\rangle, \\
  F_1&=\lvert\varphi_{21}\rangle\langle\varphi_{21}\rvert, &\quad
  &\lvert\varphi_{21}\rangle=0.500028\lvert0\rangle+(- 0.062398 + 0.498087i)\lvert1\rangle+(- 0.351826 - 0.611724i)\lvert2\rangle, \\
  F_2&=\lvert\varphi_{22}\rangle\langle\varphi_{22}\rvert, &\quad
  &\lvert\varphi_{22}\rangle=0.416357\lvert0\rangle+(- 0.421270 + 0.580072i)\lvert1\rangle+(0.375055 - 0.414762i)\lvert2\rangle, \\
  F_3&=\lvert\varphi_{23}\rangle\langle\varphi_{23}\rvert, &\quad
  &\lvert\varphi_{23}\rangle=0.555120\lvert0\rangle+(- 0.275989 - 0.322007i)\lvert1\rangle+(0.606921 - 0.378986i)\lvert2\rangle, \\
  F_4&=I-\lvert\varphi_{24}\rangle\langle\varphi_{24}\rvert, &\quad
  &\lvert\varphi_{24}\rangle=0.771642\lvert0\rangle+(0.389862 + 0.263652i)\lvert1\rangle+(0.160470 - 0.396628i)\lvert2\rangle, \\
  F_5&=I-\lvert\varphi_{25}\rangle\langle\varphi_{25}\rvert, &\quad
  &\lvert\varphi_{25}\rangle=0.759855\lvert0\rangle+(0.022187 + 0.015402i)\lvert1\rangle+(0.430543 - 0.486336i)\lvert2\rangle
\end{align*}
\end{widetext}

\newpage

\bibliography{bell}

\providecommand{\K}{{\mathrm{K}}}\providecommand{\SU}{{\mathrm{SU}}}
\begin{thebibliography}{26}
\expandafter\ifx\csname natexlab\endcsname\relax\def\natexlab#1{#1}\fi
\expandafter\ifx\csname bibnamefont\endcsname\relax
  \def\bibnamefont#1{#1}\fi
\expandafter\ifx\csname bibfnamefont\endcsname\relax
  \def\bibfnamefont#1{#1}\fi
\expandafter\ifx\csname citenamefont\endcsname\relax
  \def\citenamefont#1{#1}\fi
\expandafter\ifx\csname url\endcsname\relax
  \def\url#1{\texttt{#1}}\fi
\expandafter\ifx\csname urlprefix\endcsname\relax\def\urlprefix{URL }\fi
\providecommand{\bibinfo}[2]{#2}
\providecommand{\eprint}[2][]{\url{#2}}

\bibitem[{\citenamefont{Werner and Wolf}(2001{\natexlab{a}})}]{WerWol-QIC01}
\bibinfo{author}{\bibfnamefont{R.~F.} \bibnamefont{Werner}} \bibnamefont{and}
  \bibinfo{author}{\bibfnamefont{M.~M.} \bibnamefont{Wolf}},
  \bibinfo{journal}{Quantum Information \& Computation}
  \textbf{\bibinfo{volume}{1}}, \bibinfo{pages}{1}
  (\bibinfo{year}{2001}{\natexlab{a}}), \bibinfo{note}{arXiv:quant-ph/0107093}.

\bibitem[{\citenamefont{Kr{\"{u}}ger and Werner}(2005)}]{KruWer-0504166}
\bibinfo{author}{\bibfnamefont{O.}~\bibnamefont{Kr{\"{u}}ger}}
  \bibnamefont{and} \bibinfo{author}{\bibfnamefont{R.~F.}
  \bibnamefont{Werner}}, \emph{\bibinfo{title}{Some open problems in quantum
  information theory}}, \bibinfo{howpublished}{arXiv:quant-ph/0504166}
  (\bibinfo{year}{2005}), \bibinfo{note}{see also
  \url{http://www.imaph.tu-bs.de/qi/problems/problems.html}}.

\bibitem[{\citenamefont{Pitowsky}(1986)}]{Pit-JMP86}
\bibinfo{author}{\bibfnamefont{I.}~\bibnamefont{Pitowsky}},
  \bibinfo{journal}{J.\ Math.\ Phys.}
  \textbf{\bibinfo{volume}{27}}, \bibinfo{pages}{1556} (\bibinfo{year}{1986}).

\bibitem[{\citenamefont{Pitowsky}(1991)}]{Pit-MP91}
\bibinfo{author}{\bibfnamefont{I.}~\bibnamefont{Pitowsky}},
  \bibinfo{journal}{Math.\ Programming} \textbf{\bibinfo{volume}{50}},
  \bibinfo{pages}{395} (\bibinfo{year}{1991}).

\bibitem[{\citenamefont{Clauser et~al.}(1969)\citenamefont{Clauser, Horne,
  Shimony, and Holt}}]{ClaHorShiHol-PRL69}
\bibinfo{author}{\bibfnamefont{J.~F.} \bibnamefont{Clauser}},
  \bibinfo{author}{\bibfnamefont{M.~A.} \bibnamefont{Horne}},
  \bibinfo{author}{\bibfnamefont{A.}~\bibnamefont{Shimony}}, \bibnamefont{and}
  \bibinfo{author}{\bibfnamefont{R.~A.} \bibnamefont{Holt}},
  \bibinfo{journal}{Phys.\ Rev.\ Lett.} \textbf{\bibinfo{volume}{23}},
  \bibinfo{pages}{880} (\bibinfo{year}{1969}).

\bibitem[{\citenamefont{Werner}(1989)}]{Wer-PRA89}
\bibinfo{author}{\bibfnamefont{R.~F.} \bibnamefont{Werner}},
  \bibinfo{journal}{Phys.\ Rev.\ A} \textbf{\bibinfo{volume}{40}},
  \bibinfo{pages}{4277} (\bibinfo{year}{1989}).

\bibitem[{\citenamefont{Collins and Gisin}(2004)}]{ColGis-JPA04}
\bibinfo{author}{\bibfnamefont{D.}~\bibnamefont{Collins}} \bibnamefont{and}
  \bibinfo{author}{\bibfnamefont{N.}~\bibnamefont{Gisin}},
  \bibinfo{journal}{J.\ Phys.\ A}
  \textbf{\bibinfo{volume}{37}}, \bibinfo{pages}{1775} (\bibinfo{year}{2004}),
  \bibinfo{note}{arXiv:quant-ph/0306129}.

\bibitem[{\citenamefont{Gisin}(2003)}]{Gis-open19}
\bibinfo{author}{\bibfnamefont{N.}~\bibnamefont{Gisin}},
  \emph{\bibinfo{title}{Stronger {B}ell inequalities for {W}erner states?}},
  \bibinfo{howpublished}{O.~Kr{\"{u}}ger and R.~F.~Werner, editors, {\emph{Some
  Open Problems in Quantum Information Theory}}, arXiv:quant-ph/0504166,
  Problem~19} (\bibinfo{year}{2003}), \bibinfo{note}{see also
  \url{http://www.imaph.tu-bs.de/qi/problems/19.html}}.

\bibitem[{\citenamefont{Avis et~al.}(2004)\citenamefont{Avis, Imai, Ito, and
  Sasaki}}]{AviImaItoSas:0404014}
\bibinfo{author}{\bibfnamefont{D.}~\bibnamefont{Avis}},
  \bibinfo{author}{\bibfnamefont{H.}~\bibnamefont{Imai}},
  \bibinfo{author}{\bibfnamefont{T.}~\bibnamefont{Ito}}, \bibnamefont{and}
  \bibinfo{author}{\bibfnamefont{Y.}~\bibnamefont{Sasaki}},
  \emph{\bibinfo{title}{Deriving tight {Bell} inequalities for 2 parties with
  many 2-valued observables from facets of cut polytopes}},
  \bibinfo{howpublished}{arXiv:quant-ph/0404014} (\bibinfo{year}{2004}).

\bibitem[{\citenamefont{Avis et~al.}(2005)\citenamefont{Avis, Imai, Ito, and
  Sasaki}}]{AviImaItoSas-JPA05}
\bibinfo{author}{\bibfnamefont{D.}~\bibnamefont{Avis}},
  \bibinfo{author}{\bibfnamefont{H.}~\bibnamefont{Imai}},
  \bibinfo{author}{\bibfnamefont{T.}~\bibnamefont{Ito}}, \bibnamefont{and}
  \bibinfo{author}{\bibfnamefont{Y.}~\bibnamefont{Sasaki}},
  \bibinfo{journal}{J.\ Phys.\ A}
  \textbf{\bibinfo{volume}{38}}, \bibinfo{pages}{10971} (\bibinfo{year}{2005}),
  \bibinfo{note}{arXiv:quant-ph/0505060}.

\bibitem[{\citenamefont{Pitowsky and Svozil}(2001)}]{PitSvo-PRA01}
\bibinfo{author}{\bibfnamefont{I.}~\bibnamefont{Pitowsky}} \bibnamefont{and}
  \bibinfo{author}{\bibfnamefont{K.}~\bibnamefont{Svozil}},
  \bibinfo{journal}{Phys.\ Rev.\ A} \textbf{\bibinfo{volume}{64}}
  (\bibinfo{year}{2001}), \bibinfo{note}{arXiv:quant-ph/0011060}.

\bibitem[{\citenamefont{{Research Group Discrete Optimization, University of
  Heidelberg}}()}]{SMAPO}
\bibinfo{author}{\bibnamefont{{Research Group Discrete Optimization, University
  of Heidelberg}}}, \emph{\bibinfo{title}{{SMAPO}---``small'' 0/1-polytopes in
  combinatorial optimization}},
  \urlprefix\url{http://www.iwr.uni-heidelberg.de/groups/comopt/software/SMAPO%
/cut/cut.html}.

\bibitem[{\citenamefont{Ito et~al.}(2006)\citenamefont{Ito, Imai, and
  Avis}}]{ItoImaAvi:bell5-www}
\bibinfo{author}{\bibfnamefont{T.}~\bibnamefont{Ito}},
  \bibinfo{author}{\bibfnamefont{H.}~\bibnamefont{Imai}}, \bibnamefont{and}
  \bibinfo{author}{\bibfnamefont{D.}~\bibnamefont{Avis}},
  \emph{\bibinfo{title}{List of {Bell} inequalities for at most 5 measurements
  per party via triangular elimination}} (\bibinfo{year}{2006}),
  \urlprefix\url{http://www-imai.is.s.u-tokyo.ac.jp/~tsuyoshi/bell/bell5.html}.

\bibitem[{\citenamefont{Konno}(1976)}]{Kon-MP76}
\bibinfo{author}{\bibfnamefont{H.}~\bibnamefont{Konno}},
  \bibinfo{journal}{Math.\ Programming} \textbf{\bibinfo{volume}{11}},
  \bibinfo{pages}{14} (\bibinfo{year}{1976}).

\bibitem[{\citenamefont{Deza and Laurent}(1997)}]{DezLau:cut97}
\bibinfo{author}{\bibfnamefont{M.~M.} \bibnamefont{Deza}} \bibnamefont{and}
  \bibinfo{author}{\bibfnamefont{M.}~\bibnamefont{Laurent}},
  \emph{\bibinfo{title}{Geometry of Cuts and Metrics}},
  vol.~\bibinfo{volume}{15} of \emph{\bibinfo{series}{Algorithms and
  Combinatorics}} (\bibinfo{publisher}{Springer}, \bibinfo{year}{1997}).

\bibitem[{\citenamefont{Horodecki et~al.}(1995)\citenamefont{Horodecki,
  Horodecki, and Horodecki}}]{HorHorHor-PLA95}
\bibinfo{author}{\bibfnamefont{R.}~\bibnamefont{Horodecki}},
  \bibinfo{author}{\bibfnamefont{P.}~\bibnamefont{Horodecki}},
  \bibnamefont{and}
  \bibinfo{author}{\bibfnamefont{M.}~\bibnamefont{Horodecki}},
  \bibinfo{journal}{Phys.\ Lett.\ A} \textbf{\bibinfo{volume}{200}},
  \bibinfo{pages}{340} (\bibinfo{year}{1995}).

\bibitem[{\citenamefont{Cleve et~al.}(2004)\citenamefont{Cleve, H{\o}yer,
  Toner, and Watrous}}]{CleHoyTonWat-CCC04}
\bibinfo{author}{\bibfnamefont{R.}~\bibnamefont{Cleve}},
  \bibinfo{author}{\bibfnamefont{P.}~\bibnamefont{H{\o}yer}},
  \bibinfo{author}{\bibfnamefont{B.}~\bibnamefont{Toner}}, \bibnamefont{and}
  \bibinfo{author}{\bibfnamefont{J.}~\bibnamefont{Watrous}}, in
  \emph{\bibinfo{booktitle}{Proceedings of 19th IEEE Annual Conference on
  Computational Complexity (CCC'04)}} (\bibinfo{year}{2004}), pp.
  \bibinfo{pages}{236--249}, \bibinfo{note}{arXiv:quant-ph/0404076}.

\bibitem[{\citenamefont{Werner and Wolf}(2001{\natexlab{b}})}]{WerWol-PRA01}
\bibinfo{author}{\bibfnamefont{R.~F.} \bibnamefont{Werner}} \bibnamefont{and}
  \bibinfo{author}{\bibfnamefont{M.~M.} \bibnamefont{Wolf}},
  \bibinfo{journal}{Phys.\ Rev.\ A} \textbf{\bibinfo{volume}{64}}
  (\bibinfo{year}{2001}{\natexlab{b}}), \bibinfo{note}{arXiv:quant-ph/0102024}.

\bibitem[{\citenamefont{Avis}()}]{Avi:lrs}
\bibinfo{author}{\bibfnamefont{D.}~\bibnamefont{Avis}},
  \emph{\bibinfo{title}{lrs}},
  \urlprefix\url{http://cgm.cs.mcgill.ca/~avis/C/lrs.html}.

\bibitem[{\citenamefont{Horodecki et~al.}(2000)\citenamefont{Horodecki,
  Horodecki, and Horodecki}}]{HorHorHor-PRL00}
\bibinfo{author}{\bibfnamefont{M.}~\bibnamefont{Horodecki}},
  \bibinfo{author}{\bibfnamefont{P.}~\bibnamefont{Horodecki}},
  \bibnamefont{and}
  \bibinfo{author}{\bibfnamefont{R.}~\bibnamefont{Horodecki}},
  \bibinfo{journal}{Phys.\ Rev.\ Lett.} \textbf{\bibinfo{volume}{84}},
  \bibinfo{pages}{4260} (\bibinfo{year}{2000}).

\bibitem[{\citenamefont{Horodecki and Horodecki}(1999)}]{HorHor-PRA99}
\bibinfo{author}{\bibfnamefont{M.}~\bibnamefont{Horodecki}} \bibnamefont{and}
  \bibinfo{author}{\bibfnamefont{P.}~\bibnamefont{Horodecki}},
  \bibinfo{journal}{Phys.\ Rev.\ A} \textbf{\bibinfo{volume}{59}},
  \bibinfo{pages}{4206} (\bibinfo{year}{1999}).

\bibitem[{\citenamefont{Tsirel'son}(1987)}]{Tsi-JSovM87}
\bibinfo{author}{\bibfnamefont{B.~S.} \bibnamefont{Tsirel'son}},
  \bibinfo{journal}{J.\ Soviet Math.}
  \textbf{\bibinfo{volume}{36}}, \bibinfo{pages}{557} (\bibinfo{year}{1987}).

\bibitem[{\citenamefont{Chefles and Barnett}(1996)}]{CheBar-JPA96}
\bibinfo{author}{\bibfnamefont{A.}~\bibnamefont{Chefles}} \bibnamefont{and}
  \bibinfo{author}{\bibfnamefont{S.~M.} \bibnamefont{Barnett}},
  \bibinfo{journal}{J.\ Phys.\ A}
  \textbf{\bibinfo{volume}{29}}, \bibinfo{pages}{L237} (\bibinfo{year}{1996}).

\bibitem[{\citenamefont{Anderson et~al.}(1999)\citenamefont{Anderson, Bai,
  Bischof, Blackford, Demmel, Dongarra, Du~Croz, Greenbaum, Hammarling,
  McKenney et~al.}}]{lapack}
\bibinfo{author}{\bibfnamefont{E.}~\bibnamefont{Anderson}},
  \bibinfo{author}{\bibfnamefont{Z.}~\bibnamefont{Bai}},
  \bibinfo{author}{\bibfnamefont{C.}~\bibnamefont{Bischof}},
  \bibinfo{author}{\bibfnamefont{S.}~\bibnamefont{Blackford}},
  \bibinfo{author}{\bibfnamefont{J.}~\bibnamefont{Demmel}},
  \bibinfo{author}{\bibfnamefont{J.}~\bibnamefont{Dongarra}},
  \bibinfo{author}{\bibfnamefont{J.}~\bibnamefont{Du~Croz}},
  \bibinfo{author}{\bibfnamefont{A.}~\bibnamefont{Greenbaum}},
  \bibinfo{author}{\bibfnamefont{S.}~\bibnamefont{Hammarling}},
  \bibinfo{author}{\bibfnamefont{A.}~\bibnamefont{McKenney}},
  \bibnamefont{et~al.}, \emph{\bibinfo{title}{{LAPACK} Users' Guide}}
  (\bibinfo{publisher}{Society for Industrial and Applied Mathematics},
  \bibinfo{address}{Philadelphia, PA}, \bibinfo{year}{1999}),
  \bibinfo{edition}{3rd} ed.

\bibitem[{\citenamefont{Whaley et~al.}(2001)\citenamefont{Whaley, Petitet, and
  Dongarra}}]{WhaPetDon-PC01}
\bibinfo{author}{\bibfnamefont{R.~C.} \bibnamefont{Whaley}},
  \bibinfo{author}{\bibfnamefont{A.}~\bibnamefont{Petitet}}, \bibnamefont{and}
  \bibinfo{author}{\bibfnamefont{J.~J.} \bibnamefont{Dongarra}},
  \bibinfo{journal}{Parallel Computing} \textbf{\bibinfo{volume}{27}},
  \bibinfo{pages}{3} (\bibinfo{year}{2001}).

\bibitem[{atl()}]{atlas}
\emph{\bibinfo{title}{Automatically tuned linear algebra software {(ATLAS)}}},
  \urlprefix\url{http://math-atlas.sourceforge.net/}.

\end{thebibliography}

\end{document}